\documentclass[twocolumn,prb]{revtex4}
\usepackage{graphicx}
\usepackage{dcolumn}
\usepackage{bm}
\usepackage{amsmath}
\usepackage{amsthm}
\usepackage{amsfonts}

\def\reals{{\mathbb R}}

\begin{document}

\preprint{}
\title{Enhanced triplet superconductivity in noncentrosymmetric systems}
\author{Takehito Yokoyama, Seiichiro Onari, and Yukio Tanaka }
\affiliation{Department of Applied Physics, Nagoya University, Nagoya, 464-8603, Japan%
\\
and CREST, Japan Science and Technology Corporation (JST) Nagoya, 464-8603,
Japan }
\date{\today}

\begin{abstract}
 
 We study pairing symmetry of noncentrosymmetric superconductors based on the extended Hubbard model on square  lattice near half-filling, using the random phase approximation. We show that $d+f$-wave pairing is favored and the triplet $f$-wave state is  enhanced by Rashba type spin-orbit coupling originating from the broken inversion symmetry. The enhanced triplet superconductivity stems from  the increase of the effective interaction for the triplet pairing and the reduction of the spin susceptibility caused by the  Rashba type spin-orbit coupling which lead to the increase of the triplet component and the destruction of the singlet one, respectively. 
\end{abstract}

\pacs{PACS numbers: 74.20.Rp, 74.50.+r, 74.70.Kn}
\maketitle




%

%




Recent discovery of  heavy fermion superconductor ${\rm CePt_{3}Si}$ has opened up a new field of the study of superconductivity. \cite{Bauer} 
This is because this material does not have inversion center, which has stimulated further studies. \cite{Akazawa,Kimura,Frigeri,Yogi,Izawa,Samokhin,Bauer2,Sergienko,Bonalde,Fujimoto,Hayashi,Yokoyama,Iniotakis,Yanase}
 Because of the  broken inversion symmetry, 
Rashba type spin-orbit coupling \cite{Rashba} (RSOC) is induced, and hence 
 different parities, spin-singlet pairing and spin-triplet pairing, can be mixed in superconducting state. \cite{Gor'kov} 
From a lot of experimental and theoretical studies,
 it is believed that the most possible candidate of superconducting state in ${\rm CePt_{3}Si}$  is $s$+$p$-wave 
pairing.\cite{Frigeri,Yogi,Izawa,Samokhin,Sergienko,Bauer2,Bonalde,Fujimoto,Hayashi}
 In general, $d$+$f$-wave pairing or other mixtures are allowed in noncentrosymmetric superconductors depending on material parameters. \cite{Fujimoto}
One important  and interesting point about noncentrosymmetric superconductors is how relative magnitude of the singlet and triplet components is determined by   microscopic parameters, which has been unclear up to now.  

Spin-orbit coupling causes spin flip scattering and hence could influence superconductivity. How the superconductivity is influenced seems to depend on whether it is singlet or triplet superconductivity. 
Recent experiment of the penetration depth of Li$_2$Pd$_3$B and Li$_2$Pt$_3$B clearly shows the different behavior of the penetration depths, reflecting the difference of the strengths of the RSOC.\cite{Yuan} This implies that spin-orbit coupling enhances the spin-triplet component of noncentrosymmetric superconductors. However, its physical origin has not yet been clarified. 

Motivated by these facts, in this paper, we study pairing symmetry of noncentrosymmetric superconductors based on the extended Hubbard model with the random phase approximation (RPA). As a model calculation, we focus on the square  lattice near half-filling, which is known as a prototype of strongly correlated electron systems. We show that $d+f$-wave pairing is favored and that the triplet $f$-wave state is  enhanced by the RSOC. Its physical origin is  the increase of the effective interaction for the triplet pairings and the reduction of the spin susceptibility which lead to the increase of the triplet pairing and the reduction of the singlet one, respectively. 
This could explain the finding in Ref. \cite{Yuan}.

Let us start from explaining our model. 
We consider  square lattice without inversion center. The extended Hubbard model including the RSOC can be written as 
\begin{eqnarray}
 H =  - \sum\limits_{k,\sigma} {\left( {2t\left( {\cos k_x  + \cos k_y } \right) + \mu } \right)c_{k\sigma }^\dag  } c_{k\sigma } \nonumber  \\  + \lambda \left( {\begin{array}{*{20}c}   {\sin k_y }  \\
   { - \sin k_x }  \\
   0  \\
\end{array}} \right) \cdot {\bm{\sigma }}_{s,s' } \sum\limits_{k,s,s' } {c_{ks}^\dag  } c_{k s' } \nonumber \\ 
  + U\sum\limits_k {n_{k \uparrow } } n_{k \downarrow }  + \sum\limits_{k,\sigma ,\sigma'  } {V(k)n_{k\sigma } } n_{k\sigma'  }  
\end{eqnarray}
with $V(k) = 2V\left( {\cos k_x  + \cos k_y } \right)$. Here, $k$ represents two dimensional vector. We set lattice constant to be unity. The first term is the dispersion relation  and the second term represents the RSOC with coupling constant $\lambda$. The third and forth terms represent on-site and nearest neighbor electron-electron repulsions, respectively. 

Then, the bare Green's functions have the following form in the 2$\times$2 spin space:
\begin{eqnarray}
 \left( {\begin{array}{*{20}c}
   {G_{ \uparrow  \uparrow } \left( {k,i\omega _n } \right)} & {G_{ \uparrow  \downarrow } \left( {k,i\omega _n } \right)}  \\
   {G_{ \downarrow  \uparrow } \left( {k,i\omega _n } \right)} & {G_{ \downarrow  \downarrow } \left( {k,i\omega _n } \right)}  \\
\end{array}} \right)
 = G_ +  \left( {k,i\omega _n } \right) \\ \nonumber
  + \frac{1}{{\sqrt {\sin ^2 k_x  + \sin ^2 k_y } }}\left( {\begin{array}{*{20}c}
   {\sin k_y }  \\
   { - \sin k_x }  \\
   0  \\
\end{array}} \right) \cdot {\bm{\sigma }}G_ -  \left( {k,i\omega _n } \right),  \\
 G_ \pm  \left( {k,i\omega _n } \right) = \frac{1}{2}\left( {\frac{1}{{i\omega _n  - \xi _ +  }} \pm \frac{1}{{i\omega _n  - \xi _ -  }}} \right), \\ 
 \xi _ \pm   =  - 2t\left( {\cos k_x  + \cos k_y } \right) - \mu  \pm \lambda \sqrt {\sin ^2 k_x  + \sin ^2 k_y }   \label{xi}
\end{eqnarray}
with Matsubara frequency $\omega _n$.

The linearized $\acute{{\rm E}}$liashberg's equations with RPA in the weak coupling approximation are described as 
\begin{widetext}
\begin{eqnarray}
 \Lambda \Delta _{ss} \left( k \right) = \frac{1}{{\beta N}}\sum\limits_{q,\omega _m ,\sigma ,\sigma ' } (G_{s\sigma } \left( {q,i\omega _m } \right)G_{s\sigma ' } \left( { - q, - i\omega _m } \right)\left( { - \Gamma _{ss} \left( {k - q} \right)} \right)
\nonumber \\
 + G_{ - s\sigma } \left( {q,i\omega _m } \right)G_{ - s\sigma ' } \left( { - q, - i\omega _m } \right) U^2 \chi _{lad}^{ s,-s } \left( {k + q} \right))\Delta _{\sigma \sigma ' } \left( q \right),  \\
 \Lambda \Delta _{s, - s} \left( k \right) = \frac{1}{{\beta N}}\sum\limits_{q,\omega _m ,\sigma ,\sigma ' } {(G_{s\sigma } \left( {q,i\omega _m } \right)G_{ - s\sigma ' } \left( { - q, - i\omega _m } \right)\left( { - \Gamma _{s, - s} \left( {k - q} \right) + U^2 \chi _{lad}^{ss} \left( {k + q} \right)} \right))} \Delta _{\sigma \sigma ' } \left( q \right), \\ 
  - \Gamma _{s ,s } \left( k \right) =  - V(k) + \frac{1}{2}\left( {U + 2V(k)} \right)^2 \chi _C \left( k \right) + \frac{1}{2}U^2 \chi _S \left( k \right), \label{gammauu} \\ 
  - \Gamma _{s , - s } \left( k \right) =  - \left( {U + V(k)} \right) + \frac{1}{2}\left( {U + 2V(k)} \right)^2 \chi _C \left( k \right) - \frac{1}{2}U^2 \chi _S \left( k \right) - U\left( {U + 2V(k)} \right){\mathop{\rm Im}\nolimits} \chi ^{s , -s } \left( k \right) \label{gamma}
\end{eqnarray}
\end{widetext}
with $s=\uparrow,  \downarrow$ and inverse temperature $\beta$. 
Here, $\chi _S$ and $\chi _C$ are spin and charge susceptibilities at $\omega _n=0$, respectively, which are obtained by  
 $\chi _S  = \chi ^{ \uparrow  \uparrow }  - {\mathop{\rm Re}\nolimits} \chi ^{ \uparrow  \downarrow }$ and 
$ \chi _C  = \chi ^{ \uparrow  \uparrow }  + {\mathop{\rm Re}\nolimits} \chi ^{ \uparrow  \downarrow }  $. Note that 
 $\chi ^{ \uparrow  \uparrow }  = \chi ^{ \downarrow  \downarrow }$ and 
 $\chi ^{ \uparrow  \downarrow }  = \left( {\chi ^{ \downarrow  \uparrow } } \right)^*$ are satisfied. 
$\chi ^{ \uparrow  \uparrow }$ and $\chi ^{ \uparrow  \downarrow }$ are given by 
\begin{eqnarray}
 \left( {\begin{array}{*{20}c}
   {\chi ^{ \uparrow  \uparrow } \left( {k } \right)}  \\
   {\chi ^{ \uparrow  \downarrow } \left( {k } \right)}  \\
\end{array}} \right) = \frac{1}{A}\left( {\begin{array}{*{20}c}
   {\chi _1  + V\left( k \right)\left( {\chi _1 ^2  - \left| {\chi _2 } \right|^2 } \right)}  \\
   {\chi _2  - \left( {U + V\left( k \right)} \right)\left( {\chi _1 ^2  - \left| {\chi _2 } \right|^2 } \right)}  \\
\end{array}} \right), 
\end{eqnarray}
$A = 1 + 2V\chi _1  - U(U + V\left( k \right))\left( {\chi _1 ^2  - \left| {\chi _2 } \right|^2 } \right) 
+ 2(U + V\left( k \right)){\mathop{\rm Re}\nolimits} \chi _2 $,
\begin{eqnarray}
 \chi _1 (k)=  - \frac{1}{{\beta N}}\sum\limits_{q,\omega _n} {G_ +  \left( {k + q,i\omega _n  } \right)G_ +  \left( {q,i\omega _n } \right)},  \\ 
 \chi _2 (k) =  - \frac{1}{{\beta N}}\sum\limits_{q,\omega _n } {G_{ \uparrow  \downarrow } \left( {k + q,i\omega _n } \right)G_{ \downarrow  \uparrow } \left( {q,i\omega _n } \right)}  .
\end{eqnarray}
$ \chi _{lad}^{ \uparrow  \uparrow }$ and $\chi _{lad}^{ \uparrow  \downarrow } $ are defined as 
\begin{eqnarray}
\left( {\begin{array}{*{20}c}
   \chi _{lad}^{ \uparrow  \uparrow } (k)  \\
   \chi _{lad}^{ \uparrow  \downarrow } (k)  \\
\end{array}} \right) = \frac{1}{B}\left( {\begin{array}{*{20}c}
   { - \chi _1  + U\left( {\chi _1 ^2  - \left| {\chi _2 } \right|^2 } \right)}  \\
   { - \chi _2 }  \\
\end{array}} \right)
\end{eqnarray}
$B=\left( {1 - U\chi _1 } \right)^2  - \left| {U\chi _2 } \right|^2 $.
Notice that $\chi _{lad}^{ \uparrow  \uparrow }  = \chi _{lad}^{ \downarrow  \downarrow }$ and $  \chi _{lad}^{ \uparrow  \downarrow }  = \left( {\chi _{lad}^{ \downarrow  \uparrow } } \right)^* $ are satisfied.

In the RPA, we take into account the contributions from bubble and ladder types of diagrams. \cite{Shimahara} In the above, $\chi ^{ \uparrow  \uparrow }$ and $\chi ^{ \uparrow  \downarrow }$ stem from the bubble type of diagrams, while $\chi _{lad}^{ \uparrow  \uparrow }$  and $\chi _{lad}^{ \uparrow  \downarrow }$  originate from the ladder type of diagrams. 
Here, we ignore the nearest neighbor electron-electron repulsion in the ladder diagram for simplicity.
However, we can grasp the essence of the physics by the RPA. \cite{Miyake,Scalapino} 

By solving the $\acute{{\rm E}}$liashberg's equations, we can obtain the gap functions. 
We define singlet  component of the pair potential  and triplet one with $S_z=0$ for a later convenience as
\begin{eqnarray}
 \Delta _s  = (\Delta _{ \uparrow  \downarrow }  - \Delta _{ \downarrow  \uparrow } )/2, \\ 
 \Delta _t  = (\Delta _{ \uparrow  \downarrow }  + \Delta _{ \downarrow  \uparrow } )/2 .
\end{eqnarray}


\begin{figure}[htb]
\begin{center}
\scalebox{0.4}{
\includegraphics[width=22.0cm,clip]{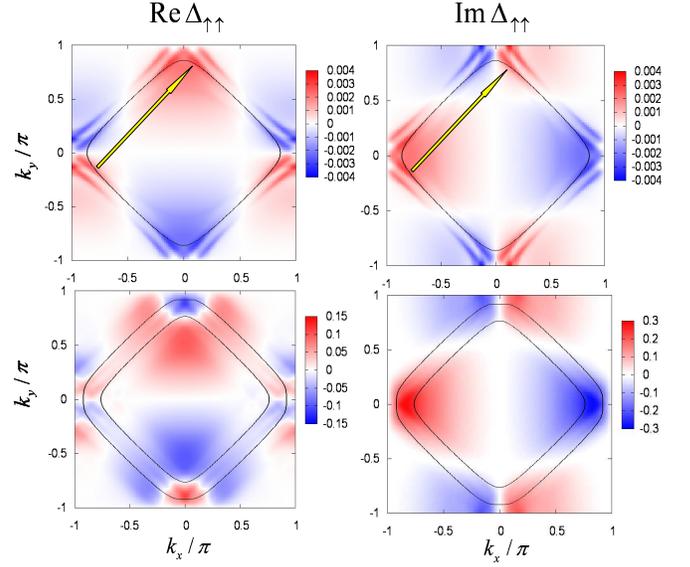}}
\end{center}
\caption{ (color) Real and imaginary parts of gap function $\Delta _{ \uparrow  \uparrow }$. We take $\lambda=0.01$ and $\lambda=0.5$ in the upper and lower figures, respectively. Solid lines represent Fermi surfaces.  Arrows indicate typical scattering processes.}
\label{f1}
\end{figure}

\begin{figure}[htb]
\begin{center}
\scalebox{0.4}{
\includegraphics[width=15.0cm,clip]{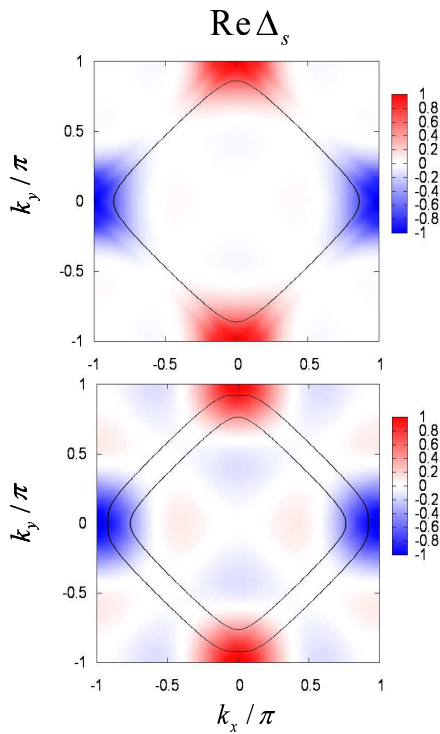}}
\end{center}
\caption{ (color)  Real part of singlet gap function $\Delta _s$. We take $\lambda=0.01$ and $\lambda=0.5$ in the upper and lower figures, respectively. Solid lines represent Fermi surfaces. }
\label{f2}
\end{figure}

\begin{figure}[htb]
\begin{center}
\scalebox{0.4}{
\includegraphics[width=22.5cm,clip]{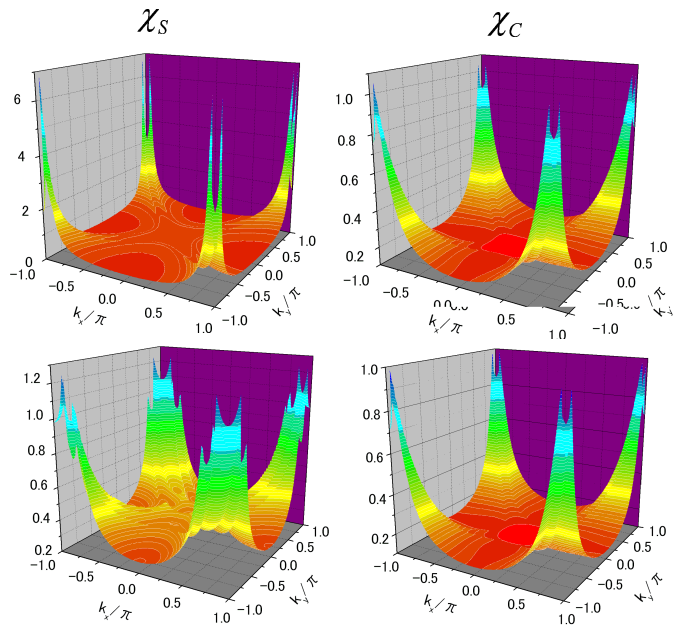}}
\end{center}
\caption{(color online) Spin and charge susceptibilities ($\chi _{S}$ and $\chi _{C}$).  We set  $\lambda=0.01$ and $\lambda=0.5$ in the upper and lower figures, respectively.}
\label{f3}
\end{figure}

\begin{figure}[tb]
\begin{center}
\scalebox{0.4}{
\includegraphics[width=15.0cm,clip]{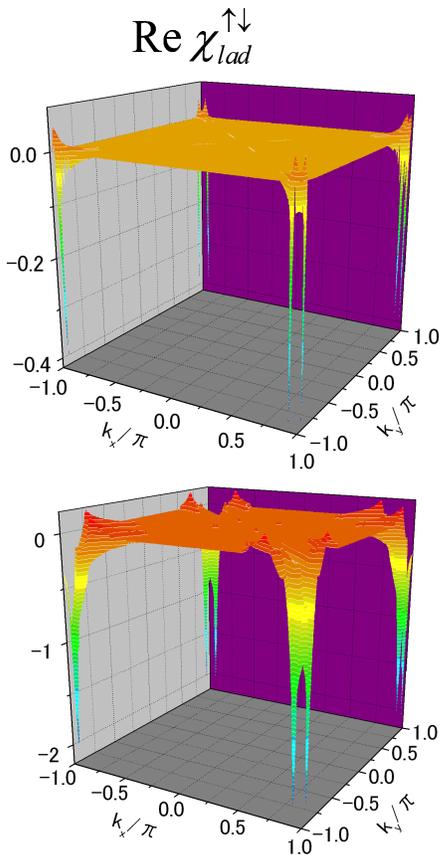}}
\end{center}
\caption{(color online) Real part of $ \chi _{lad}^{ \uparrow  \downarrow } $.  We set  $\lambda=0.01$ and $\lambda=0.5$ in the upper and lower figures, respectively.}
\label{f4}
\end{figure}

\begin{figure}[tb]
\begin{center}
\scalebox{0.4}{
\includegraphics[width=15.0cm,clip]{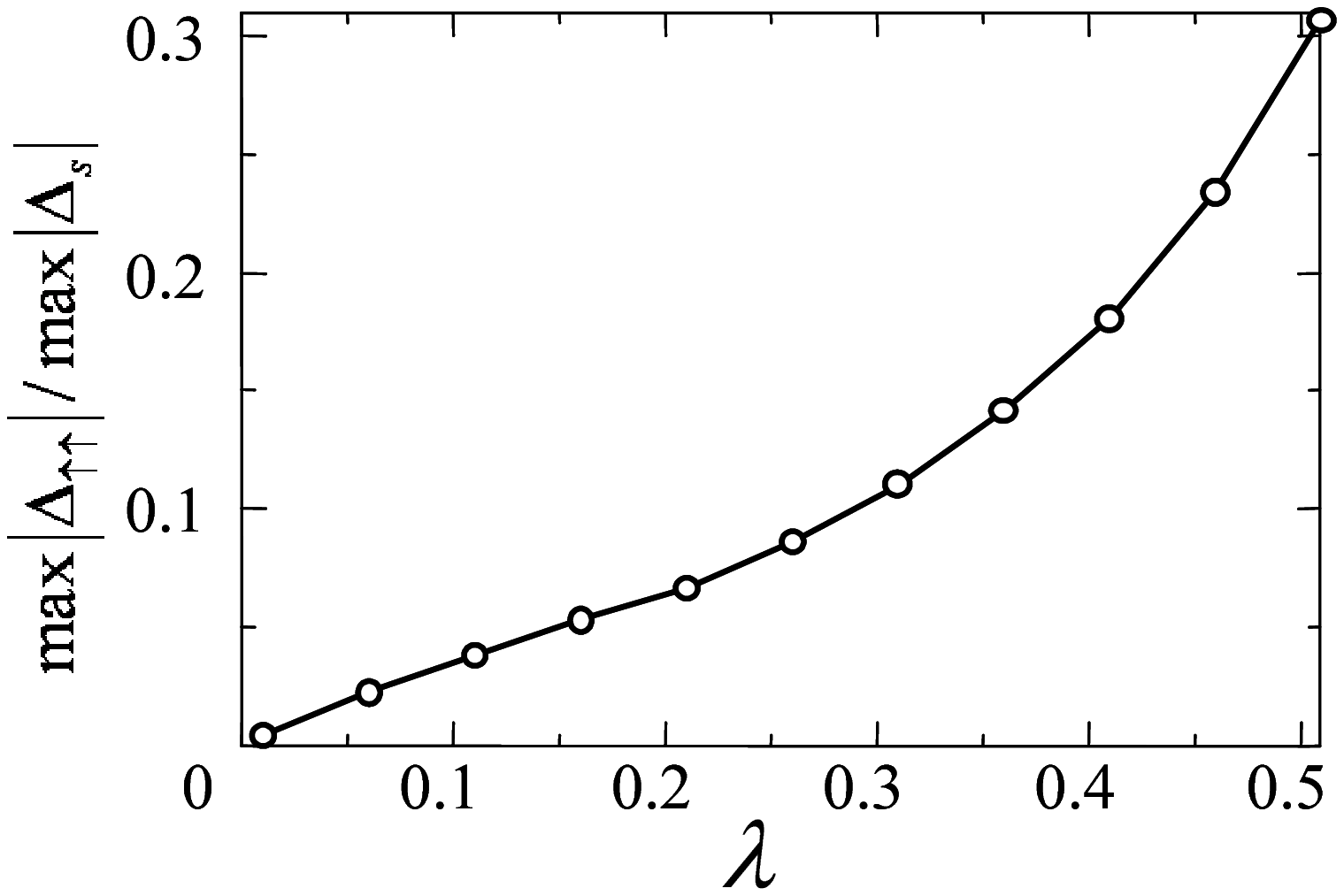}}
\end{center}
\caption{ Relative magnitude of $\max \left| {\Delta _{ \uparrow  \uparrow } } \right|$ and $\max \left| {\Delta _s } \right|$ as a function of the strength of RSOC $\lambda$.}
\label{f5}
\end{figure}

Let us first discuss the pairing symmetries. 
We put $t=1$ and set parameters as $T=0.01, N=256\times256, U=1.6, n=0.9$, and $V=0.3$, while we change the strength of RSOC $\lambda$. Here, $T, N,$ and $n$ denote the temperature, k-point meshes and the band filling, respectively. 
For these parameters, the eigenvalue $\Lambda$ is around 
$\Lambda = 0.3 \sim 0.7$.  
Gap functions are normalized by the maximum value of $\left| {\Delta _{ \uparrow  \downarrow } } \right|$. 
Note that  $\Delta _{ \downarrow  \downarrow }  =  - \Delta^* _{ \uparrow  \uparrow } $ and $\Delta _s \in \reals$ are satisfied due to the time reversal symmetry. We have found that the magnitude of the triplet component with $S_z=0$, $\Delta _t $, is negligibly small. Then the $d$-vector has no $z$-component. This is consistent with the prediction that the $d$-vector tends to be parallel to $(\sin k_y, -\sin k_x,0)$.\cite{Frigeri} 
We have also confirmed that the obtained results do not change qualitatively for other set of parameters. 

Figure \ref{f1} depicts real and imaginary parts of gap function $\Delta _{ \uparrow  \uparrow }$. As shown in this figure, it has a $f$-wave  symmetry.  We take $\lambda=0.01$ and $\lambda=0.5$ in the upper and lower figures, respectively. We can see that the magnitude of the gap function is enhanced with the increase of RSOC.

We show the real part of singlet gap function $\Delta _s$ in Fig. \ref{f2}. We also take $\lambda=0.01$ and $\lambda=0.5$ in the upper and lower figures, respectively. As can be seen from  this figure, it has a $d$-wave symmetry, which is consistent with the previous works.\cite{Miyake,Scalapino}  We can find that the symmetry  of the gap function is independent of the  RSOC. 

The appearance of $f$-wave symmetry in $\Delta _{ \uparrow  \uparrow }$ at $\lambda=0.01$  can be understood by the structures of the spin and charge susceptibilities ($\chi _{S}$ and $\chi _{C}$). They have peaks near $(\pm\pi,\pm\pi)$ as shown in  Fig. \ref{f3}. According to Eq.(\ref{gammauu}), the gap functions tend to have the same sign during the scattering process. Therefore, $f$-wave symmetry is favored. We show  typical scattering processes toward $(\pi,\pi)$ by yellow arrows in Fig. \ref{f1}. A rather complicated structure of the real part of  $\Delta _{ \uparrow  \uparrow }$ at $\lambda=0.5$ in Fig. \ref{f1} stems from the fact that the off-diagonal components of the Green's functions disappear at the van Hove singularities (see Eq.(\ref{xi})). Since the contributions from the diagonal (off-diagonal)  components are dominant near (far away from)  the van Hove singularities, we can expect line nodes in the intermediate regions. 

To understand the origin of the enhancement of the triplet pairing, we  study $\chi _{S}$ and $\chi _{C}$,  and the real part of $\chi _{lad}^{ \uparrow  \downarrow }$, which are plotted  in Fig. \ref{f3} and Fig. \ref{f4}, respectively. 
Clearly, $\chi _{S}$ is reduced by RSOC while $\chi _{C}$ is almost independent of it. This can be intuitively interpreted as follows. Spin-orbit coupling causes spin flip process and hence breaks magnetic fluctuation. On the other hand, spin flip scattering does not affect  charge fluctuation. Therefore,  $\chi _{S}$ depends on the RSOC while $\chi _{C}$ is almost independent of it. 
Since the positions of the peaks in $\chi _{S}$  and $\chi _{C}$ are almost the same, they compete with each other (see Eq.(\ref{gamma})). It is known that in such a situation, the decrease of  $\chi _{S}$ leads to the reduction of singlet pairings and hence triplet pairings could dominate.\cite{Onari} Note that $\chi _{lad}^{ \uparrow  \uparrow }$ is also reduced by the RSOC. 
Additionally as shown in Fig. \ref{f4},  the real part of $\chi _{lad}^{ \uparrow  \downarrow }$, which contributes to the effective interaction for the triplet pairings,  is enhanced by the RSOC. This results in the enhancement of the triplet component. Note that the imaginary part of $\chi _{lad}^{ \uparrow  \downarrow }$  is negligibly small. 
 These are the reasons of the enhancement of the triplet pairing by the RSOC. 

Finally, let us discuss the relative magnitude of $\max \left| \Delta _{ \uparrow  \uparrow} \right|$ and $\max \left| {\Delta _{ s } } \right|$ in detail.  In  Fig. \ref{f5},  we plot $\max \left| {\Delta _{ \uparrow  \uparrow } } \right|/\max \left| {\Delta _{ s } } \right|$ as a function of the strength of RSOC $\lambda$. It increases monotonically with $\lambda$. For a sufficiently large magnitude of $\lambda$, the magnitude of the triplet pairing could be comparable with the single one.

In conclusion, we studied pairing symmetry of noncentrosymmetric superconductors, where we used the extended Hubbard model on square lattice near half-filling with the RPA.  We found that $d+f$-wave pairing is favored and that the triplet $f$-wave state is enhanced by the RSOC. This stems from the increase of the effective interaction for the triplet pairing and the reduction of the spin susceptibility by the RSOC, resulting in the increase of the triplet pairing and the destruction of the singlet one, respectively. 

Our results are consistent with the recent experiment of the penetration depth of Li$_2$Pd$_3$B and Li$_2$Pt$_3$B which indicates that spin-orbit coupling enhances the spin-triplet component of noncentrosymmetric superconductors.\cite{Yuan} 

%
T. Y. acknowledges support by JSPS. 
%


\end{document}